\documentstyle[12pt]{article}
\begin{document}
\pagestyle{plain}
\thispagestyle{empty}
\setlength{\baselineskip} {2.5ex}
\begin {center}
{\bf Presented at the Workshop on Chiral Dynamics,\\
Massachusetts Institute of Technology, July 1994\\
Tel Aviv U. Preprint TAUP 2207-94\\
Bulletin Board: hep-ph@xxx.lanl.gov/9410215}
\end{center}
\vspace{1.0cm}
\begin{center}
{\bf \Large Pion Polarizability, Radiative Transitions, \\and Quark Gluon
Plasma Signatures\\}
\end {center}
\vspace{1.0cm}
\begin {center}
{\bf
Murray A. Moinester \\
Raymond and Beverly Sackler Faculty of Exact
Sciences,\\
School of Physics,
Tel Aviv University, 69978 Ramat Aviv, Israel\\
E-mail: murray@tauphy.tau.ac.il}
\end {center}
\vspace{1.5 cm}
\begin{center}
                            {\bf  Abstract}
\end{center}
Can one expect gamma rays rates from the QGP to be higher than from the hot
hadronic gas phase? Xiong, Shuryak, Brown (XSB) calculate photon production
from a hot hadronic gas via the reaction $\pi^- + \rho^0 \rightarrow \pi^- +
\gamma$. They assume that this reaction proceeds through the a$_1$(1260). They
also use their estimated a$_1$ radiative width to calculate the pion
polarizability; following a connection noted previously by Holstein. However,
for a$_1$(1260) $\rightarrow \pi \gamma$, the experimental radiative width is
more than two times less than the value 1.4 MeV estimated and used by XSB. We
describe how the gamma production reaction can be studied in Fermilab E781
(via the inverse reaction, with detailed balance) via the Primakoff reaction
$\pi^- + \gamma \rightarrow \pi^- + \rho^0$. Such a study can provide the data
base for evaluations of  the utility of gamma production experiments in QGP
searches. One can experimentally check the a$_1$ dominance assumption of XSB.
A remeasurement of the a$_1(1260)$ radiative width and of the pion
polarizability in Fermilab E781 will also allow us to reevaluate the
consistency of their expected relationship.

\newpage
Chakrabarty et al. \cite{chak} studied the expected gamma ray yields from hot
hadronic gases and the QGP. They suggested that gamma rays between 2-3 GeV
from the QGP outshine those from the hot hadronic gas phase. One expects many
gamma rays from QGP processes, such as a quark-antiquark annihilation $q
\bar{q}\rightarrow g \gamma$ or Compton processes such as $q g \rightarrow q
\gamma$ and $\bar{q} g \rightarrow \bar{q} \gamma$. Xiong, Shuryak, Brown
(XSB) \cite{xsb} studied gamma-ray production from a hot hadronic gas. They
calculate photon production (above 0.7 GeV) via the reaction $\pi^- + \rho^0
\rightarrow \pi^- + \gamma$. They assume that this reaction proceeds through
the a$_1$(1260). A more recent photon production calculation also involving
the
a$_1$ resonance was given by Song \cite{song}. In a hadronic gas at
high temperature, the $\pi\rho$ interaction can be near the a$_1$ resonance
\cite{xsb}. One must consider also that certain properties (masses, sizes,
parity mixing) of the $\pi$ and $\rho$ and a$_1$ change
\cite{xsb,song,dey,asa,her}, and that their numbers increase due to the
Boltzmann factor. XSB expect an increased yield from the hot hadronic gas,
higher than estimated previously by Kapusta et al. \cite{kapu}. There are many
other theoretical studies for gamma rays from hadronic gas and QGP in the
Quark Matter conferences, and elsewhere. Some relevant articles are by
Ruuskanen \cite{ruu},  Kapusta et al. \cite{kapuqm},  Alam et al. \cite{ala},
Nadeau \cite{nad}, and Schukraft \cite{sch}.

We consider an experimental determination of the $\pi^- + \rho^0 \rightarrow
\pi^- + \gamma$ total reaction rate for photon production above 0.7 GeV. The
gamma production reaction can be studied (via the inverse reaction, with
detailed balance) via the Primakoff reaction $\pi^- + \gamma \rightarrow \pi^-
+ \rho^0$. Here, the reaction takes place when an incident high energy pion
interacts with a virtual photon in the Coulomb field of a target nucleus of
atomic number Z. Such a study measures the reaction rate for normal mass pion
and rho and intermediate resonances at normal temperatures. It therefore
experimentally provides the data base for evaluations of  the utility of gamma
production experiments in QGP searches. One can experimentally check the
a$_1$ dominance assumption of XSB. The invariant mass of
the produced $\pi\rho$ system is a signature for the reaction
mechanism. For the case of $\pi\rho$ detection, one may expect the invariant
mass to show a spectrum of resonances that have a $\pi\rho$ decay branch.
These include the a$_1$(1260), $\pi$(1300), a$_2$(1320), a$_1$(1550), etc.

   High energy pion experiments at FNAL E781 \cite{russ,moin}
and CERN can obtain
new high statistics data for radiative transitions leading from the pion to
the to the a$_1$(1260), and to other resonances; via detection of the
$\pi^- \rho^0$ final state.
These radiative transition
widths are predicted by vector dominance and quark models. They were studied
in the past by different groups, but independent data would still be of value.
For a$_1$(1260) $\rightarrow \pi \gamma$, the experimental
radiative width given
\cite{ziel} is $\Gamma = 0.64 \pm 0.25 $ MeV; more than two times less than
the value 1.4 MeV estimated by XSB. It is with the 1.4 MeV width that XSB
calculate the high energy photon production cross section.

XSB also use their estimated a$_1$ radiative width to calculate the pion
polarizability, obtaining $\bar{\alpha_{\pi}} =
1.8 \times 10^{-43}$ cm$^3$. The connection
between this width and the polarizability was previously noted by Holstein
\cite {hols}. He showed that meson exchange via a pole diagram invoving the
a$_1$ resonance provides the main contribution (2.6 $\times$ 10$^{-43}$
cm$^3$) to the polarizability.

The Fermilab E781 experiment can obtain data for pion polarizability and the
a$_1$ radiative transition \cite {moin}.
A remeasurement of the a$_1(1260)$ width and of
the pion polarizability will allow us to reevaluate the consistency of their
expected relationship. In addition, E781 can study the reaction $\pi^- +
\rho^0 \rightarrow \pi^- + \gamma$, a background reaction of importance in
quark gluon plasma studies.
\newpage
This work was supported by the U.S.-Israel Binational Science Foundation,
Jerusalem, Israel.


\begin{thebibliography}{99}
\bibitem {chak} S. Chakrabarty et al., Phys. Rev. 46D (1992)3802;
Nucl. Phys. A544 (1992) 493.
\bibitem {xsb} L. Xiong, E. Shuryak, G. Brown, Phys. Rev. 46D (1992) 3798.
\bibitem {song} C. Song, Phys. Rev. 47C (1993) 2861.
\bibitem {dey} M. Dey et al.,Phys. Lett. 252B (1990) 620.
\bibitem {asa} M. Asakawa and M.C. Ko, Nucl. Phys. 560A (1993) 399.
\bibitem {her} M. Herrmann, B. L. Friman, W. Norenberg,
Nucl. Phys. 560A (1993) 411.
\bibitem {kapu} J. Kapusta et al.,Phys. Rev. 44D (1991) 2774;
H. Nadeau et al., Phys. Rev. 45C (1992) 3034.
\bibitem {ruu} P. V. Ruuskanen, QM91, Nucl. Phys. 544C (1992) 169C.
\bibitem {kapuqm} J. Kapusta et al., QM91, Nucl. Phys. 544C (1992) 485C; \\
J. Kapusta, QM93, Nucl. Phys. 566A (1994) 45C.
\bibitem {ala} J. Alam et al., QM91, Nucl. Phys. 544C (1992) 493C.
\bibitem {nad} H. Nadeau, Phys. Rev. 48D (1993) 3182.
\bibitem {sch} J. Schukraft, QGP Review Article,
J. Phys. G, Nucl. Part. Phys. 19 (1993) 1705.
\bibitem {russ} J. Russ, spokesman,
FNAL E781 Collaboration: Carnegie-Mellon U.,
Fermilab, U. Iowa, U. Rochester, U. Washington, Petersburg Nuclear Physics
Institute, ITEP (Moscow), IHEP (Protvino), Moscow State U., U. Sao Paulo,
Centro Brasileiro de Pesquisas Fisicas, Universidade Federale de Paraiba, IHEP
(Beijing), U. Bristol, Tel Aviv U., Max Planck Institut fur
Kernphysik-Heidelberg, Universidad Autonoma de San Luis Potosi; \newline
J. Russ: Proceedings of the CHARM2000 Workshop, Fermilab, June 1994, Eds. D.
M. Kaplan and S. Kwan, Fermilab-Conf-94/190, P. 111, (1994)
\bibitem {moin} M. A.
Moinester, Proceedings of the Conference on the Intersections Between Particle
and Nuclear Physics, Tucson, Arizona, 1991, AIP Conference Proceedings 243, P.
553, 1992, Ed. W. Van Oers;\\
M. A. Moinester, Bulletin Board: hep-ph@xxx.lanl.gov/9409463,
Tel Aviv U. Preprint 2204-94,
Contribution to Proceedings of Workshop on
Chiral Dynamics, Massachusetts Institute of Technology, July 1994,
Eds. A. Bernstein, B. Holstein
\bibitem {ziel} M. Zielinski et al., Phys. Rev. Lett. 52 (1984) 1195.
\bibitem {hols} B. R. Holstein, Comments Nucl. Part. Phys. 19 (1990) 239.
\end{thebibliography}
\end{document}